# An Intelligent QoS Algorithm for Home Networks

Wen-Jyi Hwang, Tsung-Ming Tai, Bo-Ting Pan, Tun-Yao Lou, and Yun-Jie Jhang

*Abstract*—A novel quality of service (QoS) management algorithm for home networks is presented in this letter. The algorithm is based on service prediction for intelligent QoS management. The service prediction is carried out by a general regression neural network with a profile containing the past records of the service. A novel profile updating technique is proposed so that the profile size can remain small for fast bandwidth allocation. The analytical study and experiments over real LAN reveal that the proposed algorithm provides reliable QoS management for home networks with low computational overhead.

## I. Introduction

THE coexistence of multiple wireless and wireline network resources is an inherent characteristic in a heterogeneous home network [1]. Furthermore, with the proliferation of network applications and appliances, diversified quality are usually desired for communication services. A Quality of Service (QoS) algorithm for the services well-suited for the heterogeneous nature of home networks will then be beneficial for efficient utilization of network resources.

A solution to the QoS problem in a home network may be based on approaches such as integrated services or multi-protocol label switching protocols [2]. However, bandwidth subscription may be required by these methods. It could result in large computational overhead for enforcing the subscription. A number of optimal tracking algorithms [3], [4] for networked control systems are also beneficial for the QoS management. Nevertheless, the systems are assumed to be time-invariant. Effective QoS management may then be difficult when the characteristics of networks are time-varying.

An alternative is to adopt service response prediction for QoS management. Given a bandwidth allocation, a service response prediction scheme forecasts the corresponding service quality. The QoS management is then carried out from the prediction results subject to a desired transmission quality. Although some existing service prediction techniques [5], [6] are effective, offline training is usually required. It may then be difficult to adopt the training results for home network environment with time-varying nature.

Manuscript received December 9, 2018; revised January 30, 2019; accepted January 31, 2019. Date of publication February 4, 2019; date of current version April 9, 2019. This work is supported by the Ministry of Science and Technology, Taiwan, under Grant MOST 107-2221-E-003-001-MY2. The associate editor coordinating the review of this letter and approving it for publication was A. Mellouk. *(Corresponding author: Wen-Jyi Hwang.)*

W.-J. Hwang, B.-T. Pan, T.-Y. Lou, and Y.-J. Jhang are with the Department of Computer Science and Information Engineering, National Taiwan Normal University, Taipei 117, Taiwan (e-mail: whwang@ntnu.edu.tw; kimi9235@gmail.com; g016214@hotmail.com; taco9395@gmail.com).

T.-M. Tai is with NVIDIA Corporation, Taipei 114, Taiwan (e-mail: ntai@nvidia.com).

Digital Object Identifier 10.1109/LCOMM.2019.2897567

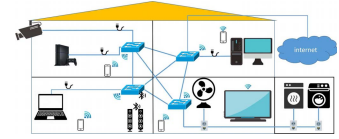

Fig. 1. An example of heterogeneous home network.

The General Regression Neural Network (GRNN) [7] has been found to be promising for service response prediction [8] without offline training. However, a drawback of the GRNN algorithm in [8] is that the computation complexities may be high for a frequently requested service. This is because the service response prediction is based on a profile consisting of past records of the service. The size of the profile grows with the service usage. This may introduce large computation overhead for QoS management.

The objective of this study is to present a novel GRNN-based QoS management algorithm for home networks. The GRNN is used for the service response prediction based on a profile of past delivery records. A novel profile updating algorithm is proposed so that the profile size of a service could remain small as the service usage grows. In this way, computation complexities may still be low for frequently used services. Analytical studies and experimental results confirm the superiority of the proposed algorithm over existing approaches for QoS management of home networks.

## II. The Proposed Algorithm

Figure 1 shows a home network example containing a number of domains. Each domain consists of a bridge, which forwards data from an appliance in that domain to the other domains. The proposed algorithm is deployed in the bridges for the QoS management. Home networks are considered in this study only as examples. The QoS management for other LAN networks such as public/office networks can also be implemented by the proposed algorithm for more applications.

Consider a typical delivery scenario in a home network, where a service delivering data from an apparatus in domain $s$ to an apparatus in domain $t$ is carried out. The corresponding bridges are connected by $n$ links. Let $x_j, j = 1, \ldots, n$, be the bandwidth of link $j$ allocated for the service, and $\mathbf{x} = \{x_1, \ldots, x_n\}$. Let $|\mathbf{x}| = \sum_{j=1}^n x_j$ be the total bandwidth of the bandwidth allocation $\mathbf{x}$. Let $R$ be the source data rate of the service. When $|\mathbf{x}| \geq R$, let Residual Allocation Bandwidth (RAB) be RAB $= |\mathbf{x}| - R$. Moreover, when $|\mathbf{x}| < R$, we define Data Loss Rate (DLR) as DLR $= R - |\mathbf{x}|$. Both RAB and DLR can be measured by bridges. A single measurement, termed Extended RAB (ERAB), can then be derived as

$$\text{ERAB} = \begin{cases} \text{RAB} & \text{when } |\mathbf{x}| \geq R \\ \text{-DLR} & \text{when } |\mathbf{x}| < R. \end{cases} \quad (1)$$

Consider a set of thresholds $\{\eta_1, \ldots, \eta_{L-1}\}$ satisfying $\eta_i < \eta_j$ for $i < j$. Let interval $I_k = (\eta_{k-1}, \eta_k]$, for $k = 2, \ldots, L-1$, and $I_1 = (-\infty, \eta_1]$, $I_L = (\eta_{L-1}, \infty)$. Let a service response level $\mathbf{y}$ be a quantized ERAB, given by

$$\mathbf{y} = k \quad \text{when ERAB} \in I_k, \tag{2}$$

where $1 \leq k \leq L$, and $L$ is the number of response levels. A QoS level can be specified for a service in terms of service responses. Let $Q$ be the total number of QoS levels available for a service. An integer $a_q$ is assigned to each QoS level $q, q = 1, \ldots, Q$. When a service is delivered with QoS level $q$, its service response level is expected to be higher than (or equal to) $a_q$, where $a_i < a_j$ for $i < j$. We can then observe from (2) that a service at the service level $q$ is expected to have ERAB larger than $\eta_{(a_q-1)}$. The expected ERAB can be viewed as the additional reserved bandwidth for a service. Higher QoS levels have larger expected ERABs so that packet losses are less likely when the source data rate surges unexpectedly.

Our goal is to find an optimal bandwidth allocation $\mathbf{x}$ based on a profile $\mathcal{T} = \{\mathbf{x}_i, \mathbf{y}_i, i = 1, \ldots, p\}$ by the GRNN, where $\mathbf{x}_i$ is a bandwidth allocation, and $\mathbf{y}_i$ is its service response. Given $\mathbf{x}$ and $\mathcal{T}$, the GRNN first computes

$$\mathbf{y}^* = \frac{\sum_{i=1}^p \mathbf{y}_i W(\mathbf{x}, \mathbf{x}_i)}{\sum_{i=1}^p W(\mathbf{x}, \mathbf{x}_i)}, \tag{3}$$

where $W(\mathbf{x}, \mathbf{x}_i) = \exp(\frac{-D(\mathbf{x}, \mathbf{x}_i)}{\sigma^2})$, and $D(\mathbf{x}, \mathbf{x}_i)$ is the squared distance between $\mathbf{x}$ and $\mathbf{x}_i$. Let $\hat{\mathbf{y}}$ be the prediction of the service response, which is obtained by rounding $\mathbf{y}^*$ in (3). In the cases the rounded number is less than 1 or larger than $L$, we set $\hat{\mathbf{y}} = 1$ and $L$, respectively.

Let $B_j$ is the maximum bandwidth at the link $j$. The search space for bandwidth allocation is given by $\mathcal{B} = \{\mathbf{x} : x_j = c\Delta, 0 \leq x_j \leq B_j\}$, where $\Delta > 0$ is the search step size, $c \geq 0$ is an integer. Let $\mathcal{O}_q$ be the set of the bandwidth allocations $\mathbf{x}$ predicted to be at QoS level $q$. It can be found by $\mathcal{O}_q = \{\mathbf{x} : \hat{\mathbf{y}} \geq a_q\}$. Let $\mathbf{x}_{\mathcal{O}_q}$ be the bandwidth allocation satisfying

$$\mathbf{x}_{\mathcal{O}_q} = \min_{\mathbf{x} \in \mathcal{O}_q} |\mathbf{x}|. \tag{4}$$

The allocation $\mathbf{x}_{\mathcal{O}_q}$ is then used for the service in the proposed algorithm. Algorithm 1 summarizes the operations of the proposed algorithm. Algorithm 2 is used for the profile updating. Let $S$ be the maximum profile size. When the current profile size $p$ is less than $S$, the new bandwidth allocation $\mathbf{x}_{p+1}$ and new response $\mathbf{y}_{p+1}$ are appended to the profile, and the size $p$ is incremented by 1. When $p$ reaches $S$, the profile $\mathcal{T}$ is updated by replacing $\{\mathbf{x}_{i^*}, \mathbf{y}_{i^*}\}$ with $\{\mathbf{x}_{p+1}, \mathbf{y}_{p+1}\}$, where

$$i^* = \begin{cases} \operatorname{argmin}_{i \in \mathcal{P}} D(\mathbf{x}_{p+1}, \mathbf{x}_i) & \text{if } \mathbf{y}_{p+1} \in \mathcal{N}, \\ \operatorname{argmin}_{i \in \mathcal{N}} D(\mathbf{x}_{p+1}, \mathbf{x}_i) & \text{otherwise,} \end{cases} \tag{5}$$

and $\mathcal{N} = \{1, \ldots, a_q - 1\}$ and $\mathcal{P} = \{a_q, \ldots, L\}$ are the set of negative responses and positive responses, respectively.

### III. QoS-AWARE BANDWIDTH ALLOCATION

A major feature of the proposed QoS algorithm is that it is QoS aware. The algorithm will increase the total bandwidth to a service for delivery quality improvement for a negative response. Alternatively, a positive response will lower or maintain the total bandwidth for the resources consumption reduction. Consider a service with QoS level $q$. We first note that the set $\mathcal{O}_q$ can be rewritten as

$$\mathcal{O}_q = \{\mathbf{x} : \mathbf{y}^* \geq a_q - 1/2\}. \tag{6}$$

From (3) and (6), we see that $\mathbf{y}^*$ and $\mathcal{O}_q$ are dependent on the profile size $p$. Let $\mathcal{O}_q(p)$ be the set $\mathcal{O}_q$ with profile size $p$. Substituting (3) to (6), we obtain

$$\mathcal{O}_q(p) = \{\mathbf{x} : \frac{\sum_{i=1}^p \mathbf{y}_i W(\mathbf{x}, \mathbf{x}_i)}{\sum_{i=1}^p W(\mathbf{x}, \mathbf{x}_i)} \geq a_q - \frac{1}{2}\}. \tag{7}$$

It can be shown that (7) is equivalent to

$$\mathcal{O}_q(p) = \{\mathbf{x} : C_1 + C_2 \geq C_3\}, \tag{8}$$

where $C_1 = \sum_{u=a_q}^L \sum_{i \in \mathcal{I}_u} (u - a_q) W(\mathbf{x}, \mathbf{x}_i)$, $C_2 = \frac{1}{2} \sum_{i=1}^p W(\mathbf{x}, \mathbf{x}_i)$, $C_3 = \sum_{u=1}^{a_q-1} \sum_{i \in \mathcal{I}_u} (a_q - u) W(\mathbf{x}, \mathbf{x}_i)$, and $\mathcal{I}_u = \{i : 1 \leq i \leq p, \mathbf{y}_i = u\}$. Note that all the terms $C_1$, $C_2$ and $C_3$ in (8) are positive.

We first investigate the impact on bandwidth allocation after appending a new record $\{\mathbf{x}_{p+1}, \mathbf{y}_{p+1}\}$ to the profile $\mathcal{T}$. Two scenarios are considered separately: $\mathbf{y}_{p+1} \geq a_q$ and $\mathbf{y}_{p+1} < a_q$. In the scenario with $\mathbf{y}_{p+1} \geq a_q$, a record with positive response is received. It can be shown that

$$\mathcal{O}_q(p+1) = \{\mathbf{x} : C_1 + C_2 + (\mathbf{y}_{p+1} - a_q + \frac{1}{2}) W(\mathbf{x}, \mathbf{x}_{p+1}) \geq C_3\}. \tag{9}$$

Clearly, with $\mathbf{y}_{p+1} \geq a_q$, all the terms in (9) are positive. Consequently, by comparing (8) with (9), it follows that

$$\mathcal{O}_q(p+1) \supseteq \mathcal{O}_q(p), \quad \text{when } \mathbf{y}_{p+1} \geq a_q. \tag{10}$$

---

**Algorithm 1** Proposed QoS Management Algorithm
**Require:** initial $\mathcal{T} = \{\mathbf{x}_i, \mathbf{y}_i, i = 1, \ldots, p\}$, $S$, $q$, and $a_q$.
1: Initialize the optimal bandwidth allocation $\mathbf{x}_{\mathcal{O}_q}$ by (4).
2: **loop**
3:   **if** service required **then**
4:     Carry out the service based on $\mathbf{x}_{\mathcal{O}_q}$.
5:     Measure the ERAB defined in (1).
6:     Compute $\mathbf{y}$ from ERAB by (2).
7:     $\{\mathbf{x}_{p+1}, \mathbf{y}_{p+1}\} \leftarrow \{\mathbf{x}_{\mathcal{O}_q}, \mathbf{y}\}$.
8:     PROFILE_UPDATE($\mathcal{T}$, $\mathbf{x}_{p+1}$, $\mathbf{y}_{p+1}$, $S$, $p$, $a_q$)
9:     Compute new $\mathbf{x}_{\mathcal{O}_q}$ based on the new $\mathcal{T}$ by (4).
10:   **end if**
11: **end loop**

---

**Algorithm 2** Profile Updating Algorithm
1: **procedure** PROFILE_UPDATE($\mathcal{T}$, $\mathbf{x}_{p+1}$, $\mathbf{y}_{p+1}$, $S$, $p$, $a_q$)
2:   **if** $p < S$ **then**
3:     $\mathcal{T} \leftarrow \mathcal{T} \cup \{\mathbf{x}_{p+1}, \mathbf{y}_{p+1}\}$, $p \leftarrow p + 1$.
4:   **else**
5:     Find $i^*$ by (5), $\mathbf{x}_{i^*} \leftarrow \mathbf{x}_{p+1}$, $\mathbf{y}_{i^*} \leftarrow \mathbf{y}_{p+1}$.
6:   **end if**
7:   **return** $\mathcal{T}$, $p$.
8: **end procedure**

From (4) and (10), we get

$$|\mathbf{x}_{\mathcal{O}_q(p+1)}| \leq |\mathbf{x}_{\mathcal{O}_q(p)}|, \quad \text{when } \mathbf{y}_{p+1} \geq a_q. \tag{11}$$

Therefore, after appending a new record $\mathbf{y}_{p+1} \geq a_q$, our algorithm lowers the total bandwidth for the next service. For the scenario $\mathbf{y}_{p+1} < a_q$, it can be shown that

$$\mathcal{O}_q(p+1) = \{\mathbf{x} : C_1 + C_2 \geq (a_q - \mathbf{y}_{p+1} - \frac{1}{2}) W(\mathbf{x}, \mathbf{x}_{p+1}) + C_3\}. \tag{12}$$

Because $\mathbf{y}_{p+1} < a_q$, all terms in (12) are positive. We then conclude from (8) and (12) that $\mathcal{O}_q(p+1) \subseteq \mathcal{O}_q(p)$. Therefore,

$$|\mathbf{x}_{\mathcal{O}_q(p+1)}| \geq |\mathbf{x}_{\mathcal{O}_q(p)}|, \quad \text{when } \mathbf{y}_{p+1} < a_q. \tag{13}$$

We next consider the cases where an old record is removed from the profile. Let $\{\mathbf{x}_{i^*}, \mathbf{y}_{i^*}\}$ be the record to be deleted. When a record with positive response is removed, $\mathbf{y}_{i^*} \geq a_q$. It can then be shown that

$$\mathcal{O}_q(p-1) = \{\mathbf{x} : C_1 + C_2 \geq (\mathbf{y}_{i^*} - a_q + \frac{1}{2}) W(\mathbf{x}, \mathbf{x}_{i^*}) + C_3\}. \tag{14}$$

Because all terms in (14) are positive, it can be derived from (8) and (14) that $\mathcal{O}_q(p-1) \subseteq \mathcal{O}_q(p)$, and

$$|\mathbf{x}_{\mathcal{O}_q(p-1)}| \geq |\mathbf{x}_{\mathcal{O}_q(p)}|, \quad \text{when } \mathbf{y}_{i^*} \geq a_q. \tag{15}$$

When a negative response is removed ($\mathbf{y}_{i^*} < a_q$), we see that

$$\mathcal{O}_q(p-1) = \{\mathbf{x} : C_1 + C_2 + (a_q - \mathbf{y}_{i^*} - \frac{1}{2}) W(\mathbf{x}, \mathbf{x}_{i^*}) \geq C_3\}. \tag{16}$$

Again, all the terms are positive in (16). Therefore, we conclude from (8)(16) that $\mathcal{O}_q(p-1) \supseteq \mathcal{O}_q(p)$, and

$$|\mathbf{x}_{\mathcal{O}_q(p-1)}| \leq |\mathbf{x}_{\mathcal{O}_q(p)}|, \quad \text{when } \mathbf{y}_{i^*} < a_q. \tag{17}$$

As shown in Algorithm 2, when profile size $p < S$, its profile updating process is simply a record appending operation. Otherwise, the updating process is a record replacement operation, which can be viewed as an appending operation followed by a removal one. Therefore, from (11)(13)(15) and (17), it can be shown that Algorithm 1 is QoS-aware.

The selection of maximum and minimum profile size can be dependent on the computation complexities for bandwidth allocation, and the variation of prediction $\mathbf{y}^*$ in (3), respectively. It can then be derived from (3) that the computation time then grows linearly with the profile size $p$. Consequently, when the limitation on computation time is imposed, the maximum profile size $S$ can then be determined. One way to find the limitation is from the usage rate of the bridge responsible for the computation of bandwidth allocation. When the usage rate is above a pre-specified threshold, the available computation capacity for further extension of the profile may be small. In this case, the computation time limitation is reached.

Let $\mathbf{y}^*(p)$ be $\mathbf{y}^*$ in (3) when the profile size is $p$. Let $\Delta \mathbf{y}^* = |\mathbf{y}^*(p+1) - \mathbf{y}^*(p)|$ be the variation of $\mathbf{y}^*$ when the profile size grows from $p$ to $p+1$. It can be shown from (3) that

$$\Delta \mathbf{y}^* \leq \frac{L-1}{\sum_{i=1}^{p+1} W(\mathbf{x}, \mathbf{x}_i)}. \tag{18}$$

Because $0 < W(\mathbf{x}, \mathbf{x}_i) \leq 1$, it follows that $\sum_{i=1}^{p+1} W(\mathbf{x}, \mathbf{x}_i) \to \infty$ as $p \to \infty$. Therefore, $\Delta \mathbf{y}^* \to 0$ as $p \to \infty$. That is, we can decrease the variation of $\mathbf{y}^*$ by increasing profile size $p$. This will result in lower variation in bandwidth allocation $\mathbf{x}_{\mathcal{O}_q}$. To enforce steady bandwidth allocation, a lower bound on $\sum_{i=1}^{p} W(\mathbf{x}, \mathbf{x}_i)$ can be pre-specified. This is equivalent to imposing a lower limit on profile size.

## IV. EXPERIMENTAL RESULTS

The proposed algorithm has been deployed in a real LAN network for performance evaluation. As shown in Figure 2, there are three bridges in the LAN, namely bridges 1, 2 and 3. Two Gigabit Ethernet links are connected between bridges 1 and 2. A link aggregation technique is employed for the combination of the links. Given a service with source data rate $R$, a data distribution scheme well suited to the proposed algorithm is then employed. Let $r_j$ be the source data rate assigned to the $j$th link. The data distribution is then based on $r_j = Rx_j/|\mathbf{x}|$. The Wi-Fi is used for the connection of bridge 3 to the other two bridges. Each bridge is implemented by Raspberry Pi 3 B+. In each bridge, the Open vSwitch [9] is deployed for packet shaping and traffic queuing. The source data packets of the target service are produced by iPerf [10]. They are delivered by user datagram protocol (UDP). The maximum throughput provided by the Ethernet adapter of each bridge is 300 Mbps. Up to seven 40-Mbps services can be delivered by the proposed system.

The proposed algorithm is deployed in bridge 3 of the LAN. It first receives ERAB reports from the other bridges. From the collected information, it then computes the new bandwidth allocation, which is sent to bridge 1 for packet shaping operations by Open vSwitch. The OpenFlow protocol is used for the bandwidth information delivery. The packet shaping operations are based on token bucket algorithm, where a dedicated token bucket is assigned to each link. The token rate of a token bucket associated with a link is identical to the allocated bandwidth to that link.

The evaluation of the two-link system shown in Figure 2 for QoS management for 40 transmissions is revealed in Figure 3. In the experiment, the network system has three QoS levels (i.e., $Q = 3$) and twelve response levels (i.e., $L = 12$). We set $\{a_1, a_2, a_3\} = \{7, 9, 11\}$ and $\{\eta_1, \ldots, \eta_{11}\} = \{-11.25, -8.75, -6.25, -3.75, -1.25, 1.25, 3.75, 6.25, 8.75, 11.25, 13.75\}$ (in Mbps). The ERAB at QoS levels 1, 2 and 3 are then expected to be larger than 1.25 Mbps, 6.25 Mbps and 11.25 Mbps, respectively. The initial profile $\mathcal{T}$ contains 16 records. The maximum profile size is $S = 31$ records. That is, for the first 15 transmissions, only the record appending operations are carried out for updating the profile. The record replacement is then executed afterwards. The maximum bandwidth at link 1 and link 2 for bandwidth allocation for a single service are $B_1 = 50$ Mbps and $B_2 = 30$ Mbps, respectively.

The source data rate $R$ and the total allocated bandwidth for each QoS level are shown in Figure 3. An additional background service is also transmitted with source data rate 40 Mbps. In this way, the effectiveness of the QoS management in the presence of other data packets can be observed.

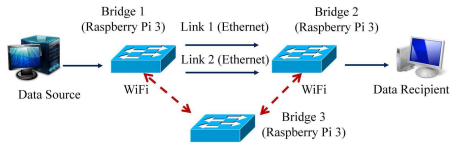

Fig. 2. A real LAN network for performance evaluation.

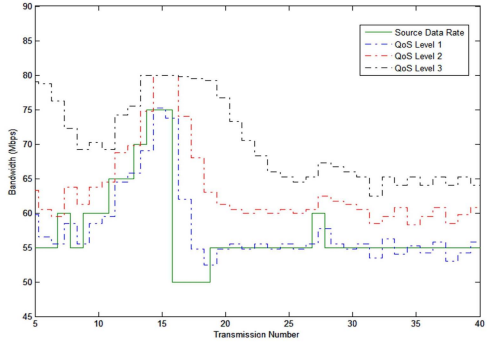

Fig. 3. The source data rate and total bandwidth allocation of the proposed algorithm at three QoS levels.

We can see from Figure 3 that the proposed algorithm provides fast tracking to $R$ at all QoS levels. This is because the proposed algorithm is QoS-aware. It is able to automatically increment or decrement the allocated bandwidth to track $R$ based on the measured ERAB. Furthermore, the measured average ERABs associated with QoS levels 1, 2 and 3 for the first 40 transmissions are 0.79 Mbps, 6.69 Mbps and 13.06 Mbps, respectively. The expected ERABs are close to the measured ones.

To further evaluate the proposed algorithm, Figure 4 reveals the total bandwidth allocation of various implementations at QoS level 2. Table I then shows the average RAB, average DLR, the computation time of bandwidth allocation at the final transmission, and the average variation of total bandwidth for two successive transmissions. It can be observed from Figure 4 that the proposed algorithm and algorithm in [8] provide similar total bandwidth allocations for all the transmissions. However, Table I shows that the algorithm in [8] has higher computation time due to larger profile size. Moreover, the algorithm in [5] based on $k$ Nearest Neighbors ($k$NN) prediction may have slower source rate tracking capability. We can also see from Table I that the average DLR and bandwidth variation of the proposed algorithm become lower as profile size $S$ increases at the expense of higher computation time. Because variation in response prediction lowers as profile $S$ increases, as shown in (18), the variation in bandwidth allocation lowers as well. The proposed algorithm then provides flexibilities on both the computational complexities and variation in bandwidth. Smaller profile sizes result in faster computation, while larger ones have more steady bandwidth allocation.

## V. CONCLUSION

Analytical studies and experiments over real home networks are provided to illustrate the effectiveness of the proposed algorithm. Analytical studies reveal that the proposed algorithm is QoS-aware. Moreover, experiments over real LAN networks exhibit that the proposed algorithm offers fast adaptation with low computational complexities to the variations of the source data rates. All the results reveal that the proposed algorithm is able to provide effective QoS management for home networks.

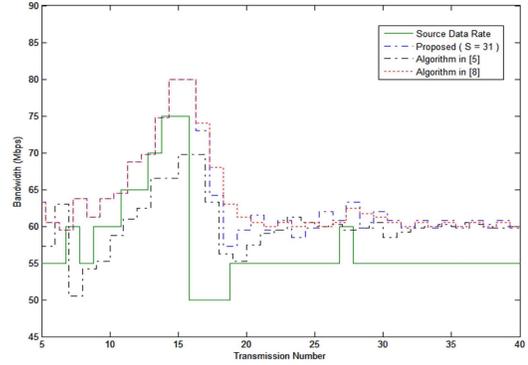

Fig. 4. Comparisons of bandwidth allocation results of various implementations at QoS level 2.

TABLE I
THE COMPARISONS OF VARIOUS IMPLEMENTATIONS ON RAB, DLR, COMPUTATION TIME AND BANDWIDTH VARIATION AT QoS LEVEL 2

|  | ave. RAB | ave. DLR | Comp. Time | ave. BW Variation |
|---|---|---|---|---|
| Proposed ($S = 16$) | 8.17 Mbps | 0.11 Mbps | 51 ms | 6.24 Mbps |
| Proposed ($S = 31$) | 6.73 Mbps | 0.04 Mbps | 85 ms | 4.68 Mbps |
| Proposed ($S = 46$) | 6.88 Mbps | 0.04 Mbps | 119 ms | 4.29 Mbps |
| Algorithm in [5] | 4.09 Mbps | 1.09 Mbps | 377 ms | 3.75 Mbps |
| Algorithm in [8] | 7.03 Mbps | 0.04 Mbps | 141 ms | 4.06 Mbps |